\newcommand{\softwarename}{\emph{ViDscribe}}

\documentclass[sigconf, pbalance=true]{acmart}

\definecolor{lightgray}{gray}{0.9}

\usepackage{xcolor}
\usepackage{subfigure}
\usepackage{subcaption}
\usepackage{balance}
\usepackage{flushend}
\definecolor{lightblue}{RGB}{244, 246, 253}
\AtBeginDocument{%
  }

\copyrightyear{2026}
\acmYear{2026}
\setcopyright{cc}
\setcctype{by}
\acmConference[CHI EA '26]{Extended Abstracts of the 2026 CHI Conference on Human Factors in Computing Systems}{April 13--17, 2026}{Barcelona, Spain}
\acmBooktitle{Extended Abstracts of the 2026 CHI Conference on Human Factors in Computing Systems (CHI EA '26), April 13--17, 2026, Barcelona, Spain}
\acmDOI{10.1145/3772363.3798744}
\acmISBN{979-8-4007-2281-3/2026/04}

\usepackage[inline]{enumitem} 

\begin{document}


\title{ViDscribe: Multimodal AI for Customizing Audio Description and Question Answering in Online Videos}

\author{Maryam Cheema}
\authornote{Equal contribution.}
\affiliation{
  \institution{Arizona State University}
  \city{Tempe}
  \state{Arizona}
  \country{USA}
}
\email{mcheema2@asu.edu}

\author{Sina Elahimanesh}
\authornotemark[1]
\affiliation{%
  \institution{Saarland University}
  \city{Saarbrücken}
  \country{Germany}
}
\email{siel00002@uni-saarland.de}

\author{Pooyan Fazli}
\affiliation{
  \institution{Arizona State University}
  \city{Tempe}
  \state{Arizona}
  \country{USA}
}
\email{pooyan@asu.edu}

\author{Hasti Seifi}
\affiliation{
  \institution{Arizona State University}
  \city{Tempe}
  \state{Arizona}
  \country{USA}
}
\email{hasti.seifi@asu.edu}


\begin{abstract}
Advances in multimodal large language models enable automatic video narration and question answering (VQA), offering scalable alternatives to labor-intensive, human-authored audio descriptions (ADs) for blind and low vision (BLV) viewers. However, prior AI-driven AD systems rarely adapt to the diverse needs and preferences of BLV individuals across videos and are typically evaluated in controlled, single-session settings. We present \softwarename, a web-based platform that integrates AI-generated ADs with six types of user customizations and a conversational VQA interface for YouTube videos. Through a longitudinal, in-the-wild study with eight BLV participants, we examine how users engage with customization and VQA features over time. Our results show sustained engagement with both features and that customized ADs improve effectiveness, enjoyment, and immersion compared to default ADs, highlighting the value of personalized, interactive video access for BLV users.
\end{abstract}




\maketitle

\section{Introduction}

Recent advances in multimodal large language models (MLLMs) enable automated audio descriptions (AD) and interactive question answering for video content \cite{li2025videoa11y,tang2025vidcomposition,avatar}, creating new opportunities to improve access for blind and low vision (BLV) users. While human-produced AD can convey rich narrative detail, capturing visual nuance, emotional tone, and complex scene dynamics \cite{Bennett2025Made}, it is costly, time-consuming, and requires specialized expertise. As a result, the vast majority of online video content remains undescribed \cite{natalie2021efficacy}. In response,  prior work suggests that AI-generated AD can provide effective and scalable access to visual media for BLV users~\cite{li2025videoa11y,Wang2021toward, describenow}.

Despite this promise, existing AI-generated AD systems largely adopt a one-size-fits-all approach and are typically evaluated in short, lab-based studies with preset videos. Recent research highlights the importance of customization \cite{describenow, jiang2024} and interactive question answering \cite{kim2023Exploring} for supporting BLV users' diverse information needs and viewing goals. However, prior customization efforts focus primarily on human-produced or edited descriptions \cite{natalie2024audio}. Little is known about how well MLLMs can support on-demand AD customization and video question answering (VQA) for online videos. Moreover, prior evaluations of AI-generated AD rarely consider longitudinal use. Consequently, we lack understanding of how BLV users’ interaction patterns, preferences, and perceptions of AI-generated descriptions evolve over time in everyday video-watching contexts. 

\begin{figure*}[t]
  \centering
  \includegraphics[width=.98\textwidth]{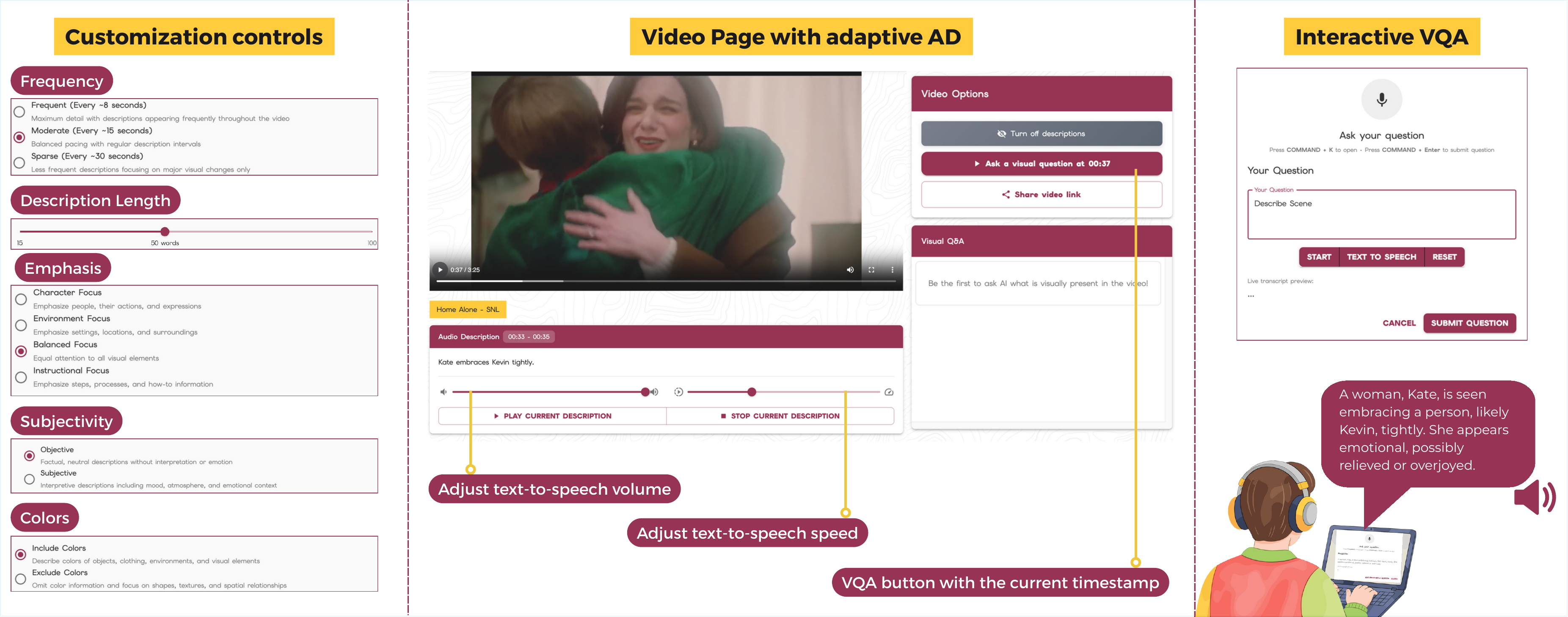}
  \caption{\softwarename\ interface, showing customization controls, video page with adaptive ADs, and interactive VQA during playback.}
  \label{fig:vidscribe}
  \Description{ViDScribe interface showing three panels. The left panel contains customization controls with sections for Frequency, Description Length, Emphasis (options: Character Focus, Environment Focus, Balanced Focus, Instructional Focus), Subjectivity (options: Objective and Subjective), and Colors (options: Include Colors and Exclude Colors). Center panel displays the video page showing a scene with two people, with audio description text "Kate embraces Kevin tightly" below. It shows controls to play the current description or stop the current description, plus sliders to adjust text-to-speech volume and speed. A Visual QA button appears at the current timestamp. The right panel shows the Interactive VQA feature with a microphone icon for voice input, a text input field for questions, and an illustration of a person with headphones using a laptop. An example VQA response is shown: ``A woman, Kate, is seen embracing a person, likely Kevin, tightly. She appears emotional, possibly relieved or overjoyed.''
}
\end{figure*}

To address these gaps, we develop \softwarename, an online platform that provides AI-generated AD, multiple forms of user-driven AD customizations, and an integrated question-and-answer feature (Figure~\ref{fig:vidscribe}). \softwarename\ supports any YouTube video, enabling BLV users to engage with content they independently select rather than researcher-curated media. Using this platform, we investigate three research questions: \textbf{(1)} How well can MLLMs support AD customization for BLV users? \textbf{(2)} What types of customizations and questions do BLV users most frequently request while watching online videos? and \textbf{(3)} How do patterns of AD customization and VQA change over time during repeated, everyday use?

To answer these questions, we conduct a longitudinal user study with eight BLV participants. Each participant completes at least five video-viewing sessions with \softwarename\ over a week. The study combines interaction logs with in-situ daily micro-surveys and post-study feedback to capture both immediate reactions and evolving usage behaviors. 
Our findings show that BLV users frequently customized ADs, reporting higher effectiveness, enjoyment, and immersion than with defaults. We report the most frequent types of VQA queries and AD customizations, as well as a shift over time toward shorter and less frequent descriptions, reflecting evolving user preferences.

This work makes two contributions: 
\begin{itemize}
    \item Design and implementation of \softwarename\footnote{\url{https://vidscribe.org/}}, an online platform that leverages MLLMs to provide automated AD, user-driven customization, and interactive VQA for arbitrary online videos; and 
    \item Empirical data from a longitudinal study that characterizes BLV individuals’ usage patterns, preferences, and perceptions of AI-generated AD customizations and VQA in everyday video-watching contexts.
\end{itemize}

\section{Related Work}

Our work lies at the intersection of interactive audio description tools for video accessibility and research on video personalization and question answering for BLV users. 

\vspace{0.1cm}
\noindent \textit{\textbf{Audio Description Tools.}}
A wide range of tools support AD creation, from manual platforms such as YouDescribe \cite{youdescribe}, Rescribe \cite{Rescribe}, and Omnidescribe \cite{OmniScribe}, to AI-based approaches including SPICA \cite{SPICA} and NarrationBot \cite{bodi2021automated,ihorn2021narrationbot}, which enable automatic generation of video descriptions \cite{campos2023Machine,Wang2021toward,li2025videoa11y,xu2025branchexplorer}. Hybrid methods combine human expertise with automation to produce timely, contextually relevant descriptions \cite{DescribePro,Natalie2023Support}. For instance, DescribePro leverages MLLMs to assist novice and professional describers in creating and collaborating on adaptive and style-aware ADs \cite{DescribePro}. While human-in-the loop tools capture rich emotional and contextual nuance \cite{yuksel2020human,yuksel2020increasing}, they cannot scale to the increasing volume of video content \cite{3playmedia}. Recent advances in large language and vision models accelerate AD generation \cite{gao-etal-2025-audio} and produce higher-quality descriptions than novice human-generated ADs \cite{li2025videoa11y,cvprw2025a11y}. Yet, most systems still produce a single, fixed description per video, limiting responsiveness to various content and individual preferences.

\vspace{0.1cm}
\noindent \textit{\textbf{Personalized ADs and Video QA.}}
BLV users have diverse needs shaped by viewing context~\cite{jiang2024, describenow, van2024shortscribe, Better2025}, and degree of blindness~\cite{chmiel2022homogenous}, highlighting the importance of personalization and interactivity in accessible video experiences \cite{VeasyGuide2025}. Prior work shows that 
professionally customized ADs can significantly improve BLV users' comprehension, satisfaction, and engagement~\cite{natalie2024audio,dcmp}, highlighting the potential for scalable, automated customization \cite{natalie2024audio}. Beyond passive descriptions, interactive question answering emerges as a complementary approach that enables users to request missing details and clarify ambiguities on demand \cite{aydin2022nonvisual,jiang2023collaborative, bodi2021automated,kulkarni2025egovitalearningplanverify}, thus enhancing agency and learning~\cite{Anderer2025Lecture}.  
Yet prior work rarely examines these features in real-world or longitudinal contexts. Building on this literature, our work investigates BLV users' week-long experience with AI-generated customized ADs combined with VQA, 
providing insights into how personalization and interactivity shape everyday video accessibility.

\section{ViDscribe}
\label{sec:platform}

\softwarename\ is a web-based, AI-powered platform designed to support BLV users through adaptive AD and interactive VQA (Figure~\ref{fig:vidscribe}). Users can watch any YouTube video by pasting its URL, and optionally selecting AD customization settings
before playback. Based on these inputs, \softwarename\ generates personalized ADs that are synchronized with the video, while allowing users to ask questions at any point during playback. Users can also adjust the speed and volume of the AD speech while watching.
\softwarename\ is implemented using a React frontend and AWS Lambda for the backend. The system leverages the \textit{Gemini 3 Pro} multimodal large language model for both AD generation and visual question answering. The interface is fully compatible with screen readers and supports keyboard-based controls for playback, customization, and AD speech-rate adjustment.

\vspace{0.1cm}
\noindent \textbf{\textit{(1) Customization Controls.}} 
To accommodate diverse preferences and viewing contexts, \softwarename\ provides six customization options informed by prior work \cite{natalie2024audio, DescribePro, describenow} and our experience with BLV users. These settings can be configured per video and adjusted as needed:
\begin{enumerate*}[label=\textbf{(\Alph*)}]
    \item \textbf{Frequency:} Controls how often audio descriptions are inserted, including frequent (every 8 seconds), moderate (15s), or sparse (30s);
    \item \textbf{Length:} Adjusts the number of words per description using a slider (15-100 words);
    \item \textbf{Emphasis:} Prioritizes the type of visual information described, such as general content, characters, instructional, or environment focus;
    \item \textbf{Subjectivity:} Switches between objective, factual descriptions and more subjective, interpretive narration;
    \item \textbf{Color Preference:} Toggles whether color attributes are explicitly described
    \item \textbf{Free-form Guidelines:} Allows users to optionally enter custom instructions via a textbox to guide AD generation further.
\end{enumerate*}
All customization settings are translated into prompt parameters to condition AD generation. \textbf{Default AD} is produced by our system when participants choose not to select any customization options. It applies 
these preset parameters: a description length of 50 words, general emphasis with objective descriptions, and color attributes enabled. 
This approach enables BLV users to access any YouTube video, including those without a human-written AD track.

\vspace{0.1cm}
\noindent \textit{\textbf{(2) Adaptive AD Generation.}} 
\softwarename\ separates AD timing from content generation to ensure temporal alignment and intelligibility. First, the AD timing module identifies suitable timestamps for inserting descriptions by following established AD guidelines and adapting the DescribePro approach \cite{DescribePro}. The module extracts the video’s audio track and visual frames, then analyzes three signals: (1) silence, (2) no-speech segments, and (3) scene changes. Silence indicates the absence of sound, while no-speech captures moments without human speech, 
helping avoid interruptions to dialogue. Scene change detection ensures that ADs align with meaningful visual transitions. The module selects timestamps where these signals overlap, prioritizing natural pauses that coincide with visual changes, and recursively splits intervals longer than the user-selected frequency to allow multiple AD insertions when appropriate.
Next, the description generation module uses \textit{Gemini 3 Pro} to generate ADs for each timestamped segment. The system provides the model with the video, the timestamps, the user’s customization settings, and a set of 42 AD guidelines from prior work~\cite{li2025videoa11y}. \textit{Gemini 3 Pro} then produces descriptions tailored to the requested frequency, length, emphasis, subjectivity, and color preferences. Full prompts and guidelines are included in Supplementary Materials. 

\vspace{0.1cm}
\noindent 
\textbf{\textit{(3) Interactive VQA Feature.}}
Beyond passive listening, \softwarename\ supports interactive visual question answering during video playback. By pressing a dedicated keyboard shortcut or on-screen button, users can ask questions (e.g., “Who just entered the room?”) either by typing or using speech-to-text. To answer a question, the system sends the user’s query along with the current video timestamp, video AD, and representative video frames to \textit{Gemini 3 Pro}, which generates a concise, context-aware response. Answers are presented as text and rendered via text-to-speech, enabling on-demand clarification and deeper engagement with visual content during everyday video watching.

\section{User Study}
We conducted a five-day study with eight BLV users to examine patterns of AD customization and VQA with \softwarename\ outside the lab. 
The study was approved by the university's ethics board. Participants received \$75 for their time.

 \vspace{0.1cm}
\noindent 
\textbf{\textit{Participants.}} We recruited eight participants (4 female, 3 male, 1 non-binary; $M_{age}=39.8$). All participants identified as blind (totally blind = 4, light perception = 3, legally blind = 1) and used screen readers. Four participants reported frequently watching videos with ADs, three occasionally, and one rarely. Seven participants had prior interaction with AI descriptions, either through user studies ($n=4$), or using visual description apps (e.g., PiccyBot, SeeingAI).

 \vspace{0.1cm}
\noindent \textbf{\textit{Study Procedure.}} The study consisted of (1) an introductory session, (2) a five-day usage period over a week, and (3) an end-of-week survey. In the introductory session (20–30 minutes), participants provided consent, completed a demographics questionnaire, and were introduced to \softwarename. Participants then used the system to watch at least 10 minutes of video per day for at least five days at times and locations of their choice. 
After each day, they completed a short survey about their experience, followed by a final survey at the end of the week.

 \vspace{0.1cm}
\noindent 
\textbf{\textit{Data Collection.}} 
In end-of-day surveys, participants rated the effectiveness, enjoyment, and immersion of watching videos with \softwarename. When used, they also rated customization alignment, helpfulness, and accuracy of VQA responses. The final survey included System Usability Scale (SUS) scores. We analyzed VQA questions using a closed-code codebook developed by one author and independently applied by two authors. For each question, both authors reviewed the video at the timestamp when the question was asked to determine whether the system's response was accurate and answerable based on the visual information. Disagreements in the coding for VQA types and accuracy were resolved through discussion. We analyzed open-ended survey responses using open coding. The system logged uploaded videos and user customizations.

\begin{figure*}[htbp]
    \centering
    \subfigure[AI description ratings (Customized vs. Default).]{
        \includegraphics[width=0.35\textwidth]{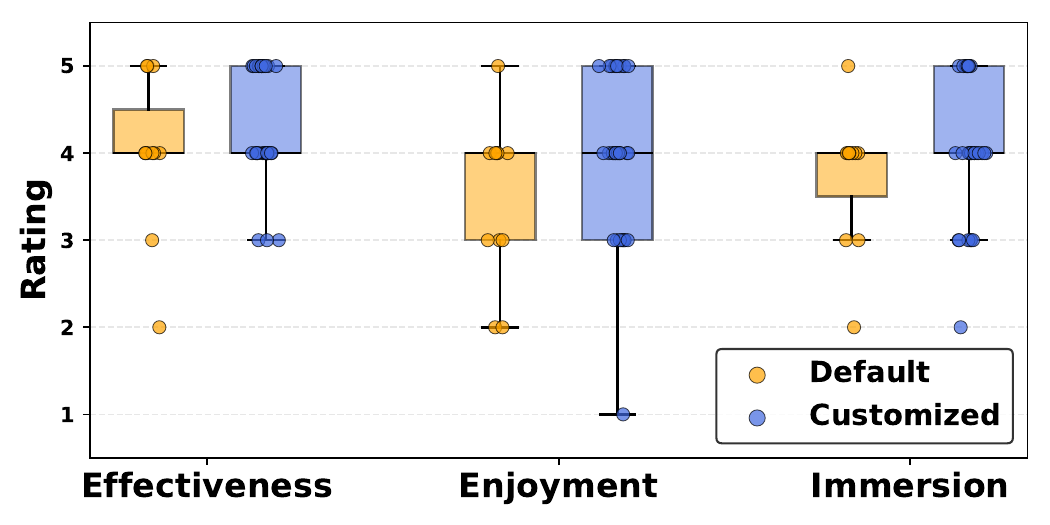}
        \label{fig:customization}
    }
     \subfigure[End of week ratings]{
       \includegraphics[width=0.61\textwidth]{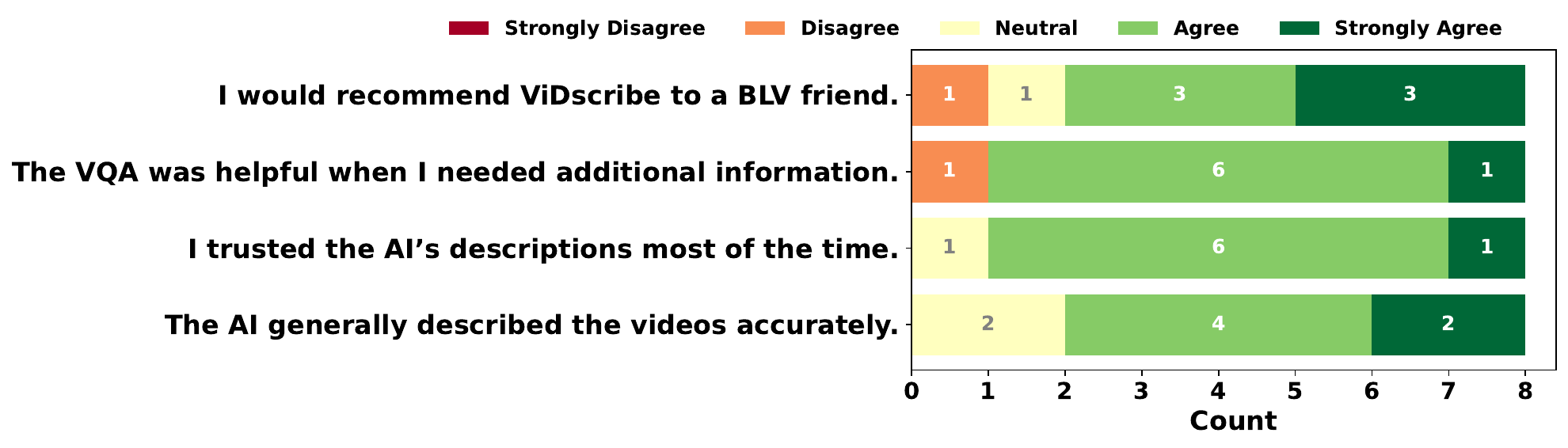}
        \label{fig:end_of_week_ratings}
    }\hspace{0.1cm}
    \caption{User ratings for (a) customized vs. default ADs in daily surveys, and (b) \softwarename\ in the end-of-week survey.} 
     \label{fig:user_experience}
\Description{Two subfigures showing user study results. (a) Horizontal stacked bar chart titled "End of week ratings" displaying four survey questions with response counts. From top to bottom: ``I would recommend ViDScribe to a BLV friend'' shows 1 Strongly Disagree, 1 Disagree, 3 Agree, 3 Strongly Agree; ``The VQA was helpful when I needed additional information'' shows 1 Disagree, 6 Agree, 1 Strongly Agree; ``I trusted the AI's descriptions most of the time'' shows 1 Disagree, 6 Agree, 1 Strongly Agree; ``The AI generally described the videos accurately'' shows 2 Neutral, 4 Agree, 2 Strongly Agree. (b) Box plots titled ``AI description ratings (Customized vs. Default)'' showing three metrics: Effectiveness, Enjoyment, and Immersion. Customized conditions show consistently higher ratings across all three metrics, with some outliers visible in the data.
}
\end{figure*}
\section{Results}
Over the longitudinal study, participants uploaded 81 videos (average duration = 3:37), 51 of which used customizations (i.e., 63\%). They completed 45 daily surveys (i.e., diary entries), including 31 customization-alignment ratings and 26 VQA ratings. Below, we report findings on their experience, customization and VQA usage, and changes over time.

\vspace{0.1cm}
\noindent \textbf{\textit{User Experience. }}
AI customizations received higher effectiveness, enjoyment, and immersion ratings than under default settings (Fig.~\ref{fig:customization}). 
Enjoyment showed the largest increase ($M_{custom}$ = 3.97; $M_{default}$ = 3.45), echoed in qualitative feedback: \textit{[P5] ``
the last video was actually quite enjoyable...I could see myself actually using this.''} Ratings for immersion($M_{custom}$ = 4.06; $M_{default}$ = 3.72) and effectiveness ($M_{custom}$ = 4.32; $M_{default}$ = 4.00) showed similar trends. Due to the small sample size and unequal numbers of observations, we did not conduct statistical significance tests.

Ratings for the VQA feature were positive but comparatively lower. Participants asked a total of 66 questions across the study. In the final survey, participants rated VQA particularly useful when seeking additional information (Fig~\ref{fig:end_of_week_ratings}).
In daily surveys, participants rated how much VQA helped them understand videos as 3.46 out of 5, while perceived accuracy was rated slightly lower at 3.27. Based on our manual analysis of VQAs, the lower ratings likely stem from cases in which participants ($n = 2$) needed to reframe or rewind videos to ask their questions, as the current VQA implementation uses only the current frame and a few adjacent frames to provide an answer. 

Overall, \softwarename\ received an average SUS score of 70.6 ($SD = 19.4$), above the average benchmark for web-based interfaces of 68~\cite{bangor2008SUS}. Participants suggested improving keyboard navigation, particularly for accessing VQA ($n=3$). Additional feedback highlighted a desire to save customization and text-to-speech settings for future use. The end-of-week ratings suggested trust in the AI-generated descriptions, and most participants ($n=6$) indicated they would recommend the system to others (Fig.~\ref{fig:end_of_week_ratings}).

\begin{table}[h]
\centering
\setlength{\tabcolsep}{4pt}
\footnotesize
\resizebox{\linewidth}{!}{\begin{tabular}{llc|llc}
\toprule
\textbf{Type} & \textbf{Option} & \textbf{n (\%)} & \textbf{Type} & \textbf{Option} & \textbf{n (\%)} \\
\toprule
\textbf{1. Emphasis} & general & 27 (52.9) &  \textbf{4.Color} & include & 41 (80.4) \\
 & character & 15 (29.4) &  & exclude & 10 (19.6) \\
 & instructional & 6 (11.8) &  \textbf{5. Subjectivity} & objective & 37 (72.5) \\
 & environment & 3 (5.9) &  & subjective & 14 (27.5) \\
 \textbf{2. Frequency} & 8 seconds & 28 (54.9) &  \textbf{6. Length} & 15--25 words & 21 (41.2) \\
 & 15 seconds & 22 (43.1) &  & 26--50 words & 25 (49.0) \\
 & 30 seconds & 1 (2.0) &  & 51--75 words & 2 (3.9) \\
 \textbf{3. Free-form Guidelines} & & 12 (23.5) & & 76--100 words & 3 (5.9)\\
\bottomrule
\end{tabular}
}
\caption{Six AD customization types and distribution of user-selected customizations.}
\label{tab:customization_counts}
\end{table}

\vspace{0.1cm}
\noindent \textbf{\textit{Types of Customizations and Questions.}}
Table~\ref{tab:customization_counts} summarizes the AD customization options selected across videos. Participants more frequently opted for general emphasis and objective descriptions. However,  character focus was popular for \emph{Sports} (3/5) and \emph{Film \& Animation} (4/14) videos. The use of free-form guidelines was also higher in \emph{Sports} (2/5) and \emph{Film \& Animation} (5/12), where many of these guidelines concerned character attributes. For example, requests were made to include character names (\textit{``Include character names''}, n=5) or to describe appearance (\textit{``Describe the apparent age and physical health of the people in the video,''} n=2). This aligns with the preference for character emphasis for \emph{Film \& Animation} genre. Compared to other genres, \emph{Film \& Animation} and \emph{Entertainment} videos showed greater use of subjective descriptions (5/14 and 3/5, respectively), suggesting varied preferences for visually rich content. Instructional emphasis was particularly popular for \emph{How-To videos} (6/11): \textit{``The instruction emphasis is very awesome for educational, crafts, and exercise videos (P7).''} Other than one instance, no videos used the 30-second frequency setting, and most videos used descriptions under 50 words, indicating a general preference for shorter, more frequent descriptions.

\begin{figure}[t]
\centering
\includegraphics[width=\linewidth]{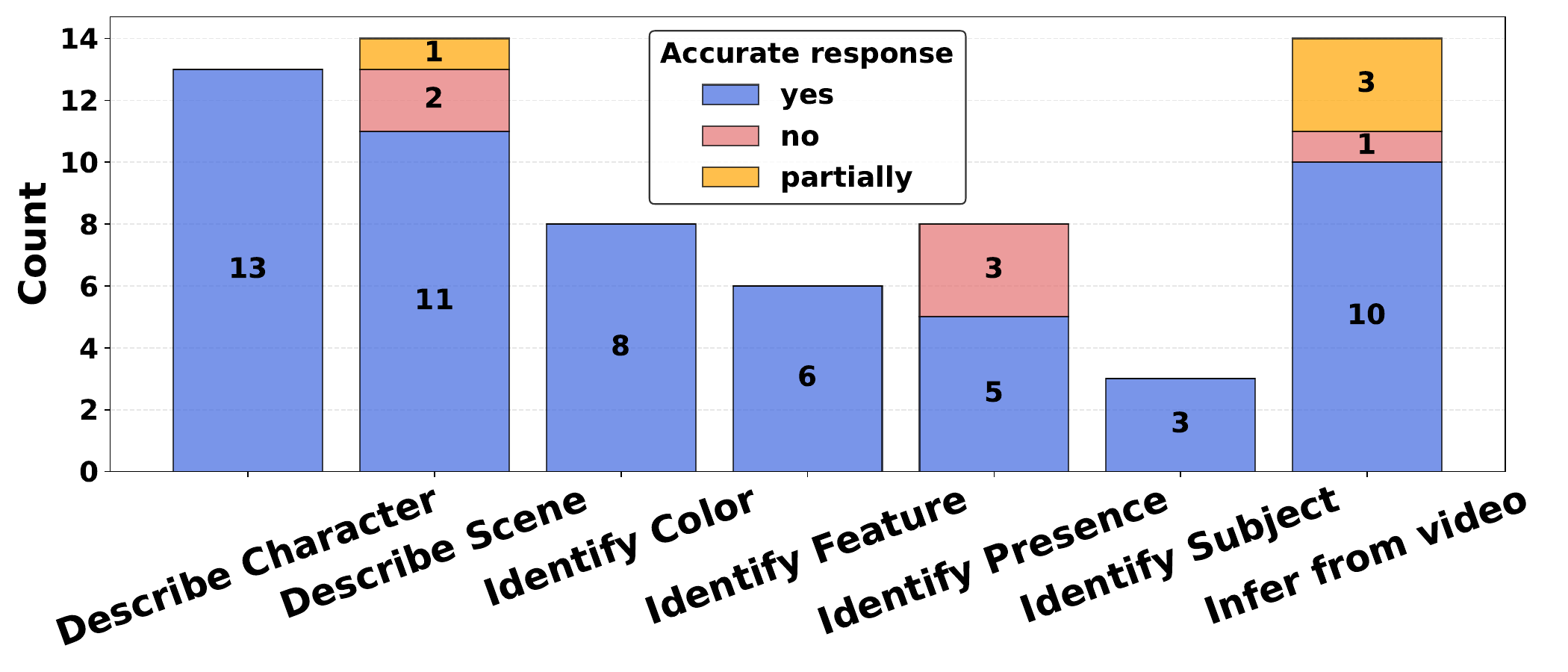}
\caption{Distribution of VQA types and response accuracy.}
\label{fig:question_types}
\Description{Bar chart showing the distribution of VQA question types and response accuracy. The x-axis lists question types (Describe Character, Describe Scene, Identify Color, Identify Feature, Identify Presence, Identify Subject, Infer from Video), and the y-axis shows counts. Bars are stacked by response accuracy, indicating that most responses across categories are accurate, with some partially accurate and incorrect responses varying by question type.}
\end{figure}

Figure~\ref{fig:question_types} shows the distribution of question types. Describing scene, character, and identifying color or presence were frequently asked questions. Several participants asked questions that required audio-visual inference (e.g., \emph{Who said ``I'm not on your team?''}), for which the VQA system sometimes failed to respond.
In addition to these common patterns, participants reported meaningful use cases. For instance, P2 used VQA to obtain product-specific information (e.g., \emph{What color is the cot?}), noting: \textit{``I actually used this video and description [VQA] since I got one of these cots for Christmas.''}

\vspace{0.1cm}
\noindent 
\textbf{\textit{Longitudinal Changes. }}
Data showed notable patterns in how participants adjusted frequency and length parameters over time. Over subsequent days, participants increasingly opted for shorter description lengths, where the average reduced from 47.7 ($SD=21.0$) to 33.3 ($SD=10.9$). This trend aligns with qualitative feedback, where in the end of week survey, four participants explicitly reported a preference for reduced length: ``\textit{I liked the mid-length description, 25 words max. (P1)}'' Participants indicated that higher word counts often introduced ``AI fluff''. A similar pattern was observed for description frequency, with participants gradually favoring longer intervals (e.g., 15 seconds) as the week progressed to reduce interruptions and repetition. In contrast, no clear patterns emerged for emphasis or subjectiveness, suggesting these choices were primarily personal preference and content-driven. 

Customization and VQA usage also remained consistent throughout the week. Specifically, perceived alignment of all customizations with the ADs remained high, with emphasis receiving the highest alignment rating ($M = 4.22$, $SD = 0.56$), followed by length ($M = 4.10$, $SD = 0.70$), frequency ($M = 4.03$, $SD = 0.75$), and subjectiveness ($M = 3.97$, $SD = 0.55$). Lower ratings for frequency and subjectiveness were due to perceived mismatches with user preferences, as P4 noted: \textit{``Sometimes it felt like the AD wasn't as frequent in the Mortal Kombat trailer. In the Kill Bill trailer, the subjective setting was too subjective.''} Together, these results suggest persistent perceptions across viewing sessions.

\section{Discussion and Conclusion}

We discuss implications of our findings for future work and outline the study's limitations. Regarding the efficacy of MLLMs in AD customization (\textbf{RQ1}), our findings suggest participants have an overall positive perception of AI customizations in line with prior findings on customization by professional describers~\cite{natalie2024audio}. Our results also suggest high trust in the generated descriptions, but as the trust in AI increases, it raises concerns about correctness and hallucinations~\cite{Li_2025_CVPR}. Prior research shows that users place high trust in longer descriptions regardless of their accuracy~\cite{huh2024long}, a pattern that may similarly apply to customized descriptions. To mitigate such risks, AI accessibility systems can incorporate human-in-the-loop mechanisms and flag videos for review based on content or uncertainty in descriptions.

With respect to types of customization and questions (\textbf{RQ2}), \emph{emphasis} emerged as the most preferred customization, with instructional emphasis favored for how-to, and character focus for entertainment content. Participants also often asked both scene-level and inferential questions. The combination of free-form guidelines and customization options in \softwarename\ offers an opportunity to develop genre-specific AD guidelines that not only prompt MLLMs but also train novice describers. Also, the prevalence of inferential questions underscores the need for VQA systems that adopt a two-step approach: first, identifying relevant video segments, and then generating contextually grounded responses.

Regarding changes over time (\textbf{RQ3}), participants consistently used customizations and VQAs throughout the week. Our results suggest that shorter and less frequent descriptions were preferred over time. One possible explanation is that as participants became more familiar with the content style and system capabilities, they learned how much visual detail they could obtain relative to the number and frequency of ADs, leading them to favor more concise descriptions that reduced cognitive load and viewing interruption, similar to how professional describers shorten AI-generated or novice-written ADs ~\cite{DescribePro}. 
Other customization choices varied by personal preference, as did the types of questions participants asked. These findings highlight a critical gap. While prior longitudinal data have examined BLV genre preferences~\cite{pitcher2024youdescribe}, and interaction with image-QA~\cite{scene2024investigating}, longitudinal engagement with AD customization and video-QA remains underexplored. Platforms like \softwarename\ enable the collection of longitudinal data at scale, by capturing users' customization choices and question-asking behavior over extended periods, generating rich behavioral data to inform adaptive systems that evolve ADs alongside users’ preferences.

\noindent \textbf{\textit{Limitations.}} Although longitudinal, the one-week study duration may not be sufficient to capture longer-term adaptation of customizations. Additionally, to reduce participant burden, we opted to collect surveys (i.e., diary entries) at the end of each day rather than after each video. Gathering insights on individual video customizations could provide statistical evidence on the efficacy of AI customizations. Finally, the study involved eight BLV participants. Future work should look into how these customizations and VQA features scale across larger user populations. 

In conclusion, \softwarename\ enhances access and provides insights into user-driven customization and VQA usage, informing the design of future interactive AD systems.

\begin{acks}
This research was supported by the National Eye Institute (NEI) of the National Institutes of Health (NIH) under award number R01EY034562. The content is solely the responsibility of the authors and does not necessarily represent the official views of the NIH. We thank all the participants for their time. We also thank Samuel Martin for his contributions to developing \softwarename, as well as to the student volunteers who created descriptions using \softwarename\ and provided valuable feedback.
\end{acks}
\bibliographystyle{ACM-Reference-Format}
\bibliography{references}

\clearpage
\appendix
\balance

\section{Appendix: Prompt Templates Used in \softwarename}
\label{app:prompts}
Below, we present the AD prompt, along with the general guidelines and customization prompts used to control emphasis, subjectivity, and color preferences. We also include the Interactive VQA prompt.
\subsection{Base Audio Description Prompt}

\setlength{\fboxsep}{8pt} 
\setlength{\fboxrule}{0.5mm}
\vspace{0.2cm}
\noindent
\fcolorbox{black}{lightblue}{ 
    \begin{minipage}{0.94\linewidth}
    You are an AI designed to assist in creating high-quality and contextually rich descriptions for videos, aimed at enhancing accessibility for blind and low-vision (BLV) users.
    The input consists of a video. Based on the video, craft audio descriptions that are highly personalized, based on guidance from the BLV.
    The input also consists of a set of timestamps. The timestamps are the start timestamps where the description will play. Ensure that the descriptions align with the visual content present at the timestamps.
    You must follow all the given instructions. You should avoid any prefatory language, such as `the video shows`. Follow the General and Customized guidelines shared by the user. You should prioritize the customized guidelines while adhering to the general Guidelines:

    GENERAL AUDIO DESCRIPTION GUIDELINES:
    
    \hspace*{1.5em} \{ -- GENERAL\_GUIDELINES -- \}

    CUSTOM GUIDELINES SPECIFIED BY USER:

    \hspace*{1.5em} - Description Length: Target approximately \{TARGET\_LENGTH \} words per description segment

    \hspace*{1.5em} - Emphasis: \{ EMPHASIS\_PROMPT \}
    
    \hspace*{1.5em} - Style: \{ SUBJECTIVITY\_PROMPT \}

    \hspace*{1.5em} - Color Descriptions: \{ COLOR\_PREFERENCES\_PROMPT \}

    \hspace*{1.5em} - User Guidelines: \{ FREE\_FORM\_GUIDELINES \}

     TIMESTAMPS:
     
     \hspace*{1.5em} \{ TIMESTAMPS \}
    
    If the timestamps are present at any point where there is speech, adjust the timestamp slightly to ensure the description plays after the description.
    
    IMPORTANT: Your response must be valid JSON matching the VideoMetadata schema.

    \hspace*{1.5em} \{ VIDEO\_FRAMES \}
\end{minipage}
}

\subsection{General Guidelines}

\setlength{\fboxsep}{8pt} 
\setlength{\fboxrule}{0.5mm}
\vspace{0.2cm}
\noindent
\fcolorbox{black}{lightblue}{ 
    \begin{minipage}{0.94\linewidth}
    1. Avoid over-describing - Do not include non-essential visual details.
    
    2. Description should not be opinionated unless content demands it.
    
    3. Choose a level of detail based on plot relevance when describing scenes.
    
    4. Description should be informative and conversational, in present tense and third-person omniscient.
    
    5. The vocabulary should reflect the predominant language/accent of the program and should be consistent with the genre and tone of the content while also mindful of the target audience. Vocabulary used should ensure accuracy, clarity, and conciseness.

    
    
    

    \end{minipage}
}

\setlength{\fboxsep}{8pt} 
\setlength{\fboxrule}{0.5mm}
\vspace{0.2cm}
\noindent
\fcolorbox{black}{lightblue}{ 
    \begin{minipage}{0.94\linewidth}

    6. Consider historical context and avoid words with negative connotations or bias.
    
    7. Pay attention to verbs - Choose vivid verbs over bland ones with adverbs.
    
    8. Use pronouns only when clear whom they refer to.
    
    9. Use comparisons for shapes and sizes with familiar and globally relevant objects.
    
    10. Maintain consistency in word choice, character qualities, and visual elements for all audio descriptions.

    11. Tone and vocabulary should match the target audience's age range.

    12. Ensure no errors in word selection, pronunciation, diction, or enunciation.
    
    13. Start with general context, then add details.
    
    14. Describe shape, size, texture, or color as appropriate to the content.
    
    15. Use first-person narrative for engagement if required to engage the audience.

    16. Use articles appropriately to introduce or refer to subjects.
    
    17. Prefer formal speech over colloquialisms, except where appropriate.
    
    18. When introducing new terms, objects, or actions, label them first, and then follow with the definitions.
    
    19. Describe objectively without personal interpretation or comment. Also, do not censor content.
    
    20. Deliver narration steadily and impersonally (but not monotonously), matching the program's tone.
    
    21. It can be important to add emotion, excitement, and lightness of touch at different points. Adjust style for emotion and mood according to the program's genre.
    
    22. If it is children's content, tailor language and pace for children's TV, considering audience feedback.
    
    23. Do not alter, filter, or exclude content. You should describe what you see. Try to seek simplicity and succinctness in your description.
    
    24. Prioritize what is relevant when describing action as to not affect user experience.
    
    25. Include location, time, and weather conditions when relevant to the scene or plot.
    
    26. Focus on key content for learning and enjoyment when creating audio descriptions. This is so that the intention of the program is conveyed.
    
    27. When describing an instructional video/content, describe the sequence of activities first.
    
    28. For a dramatic production, include elements such as style, setting, focus, period, dress, facial features, objects, and aesthetics.
    
    29. Describe what is most essential for the viewer to know in order to follow, understand, and appreciate the intended learning outcomes of the video/content.

    30. Audio description should describe characters, locations, time and circumstances, on-screen action, and on-screen information.

    31. Describe only what a sighted viewer can see.

    

    
  \end{minipage}
}  

\setlength{\fboxsep}{8pt} 
\setlength{\fboxrule}{0.5mm}
\vspace{0.2cm}
\noindent
\fcolorbox{black}{lightblue}{ 
    \begin{minipage}{0.94\linewidth}

    
    32. Describe main and key supporting characters' visual aspects relevant to identity and personality. Prioritize factual descriptions of traits like hair, skin, eyes, build, height, age, and visible disabilities. Ensure consistency and avoid singling out characters for specific traits. Use person-first language.
    
    33. If unable to confirm or if not established in the plot, do not guess or assume racial, ethnic or gender identity.
    
    34. When naming characters for the first time, aim to include a descriptor before the name (e.g., a bearded man, Jack).
    
    35. Description should convey facial expressions, body language and reactions.
    
    36. When important to the meaning/intent of content, describe race using currently-accepted terminology.
    
    37. Avoid identifying characters solely by gender expression unless it offers unique insights not apparent otherwise to visually impaired viewers.
    
    38. Describe character clothing if it enhances characterization, plot, setting, or genre enjoyment.

    39. If text on the screen is central to understanding, establish a pattern of on-screen words being read. This may include making an announcement, such as 'Words appear'.
    
    40. In the case of subtitles, the describer should read the translation after stating that a subtitle appears.
    
    41. When shot changes are critical to the understanding of the scene, indicate them by describing where the action is or where characters are present in the new shot.
    
    42. Provide description before the content rather than after.
\end{minipage}
}

\subsection{Customization Prompt: Description Subjectivity}

\setlength{\fboxsep}{8pt} 
\setlength{\fboxrule}{0.5mm}
\vspace{0.2cm}
\noindent
\fcolorbox{black}{lightblue}{ 
    \begin{minipage}{0.94\linewidth}
    
     subjectiveness\_guidelines = \{
     
        \hspace*{1.5em} 'objective': 'Maintain strict factual neutrality. Describe only observable visual elements without interpretation or emotional inference unless clearly visible. Avoid assumptions about motivations, intentions, or unstated emotional states. Use neutral, descriptive language.',
        
        \hspace*{1.5em} 'subjective': 'Use interpretive language to convey atmosphere, emotional mood, and inferred character feelings when they reasonably align with visual cues. Use expressive vocabulary to enhance immersion for the BLV user. Include mood, tone, and emotional context.'
    
        \}
\end{minipage}
}

\subsection{Customization Prompt: Color Preference}

\setlength{\fboxsep}{8pt} 
\setlength{\fboxrule}{0.5mm}
\vspace{0.2cm}
\noindent
\fcolorbox{black}{lightblue}{ 
    \begin{minipage}{0.94\linewidth}

    color\_preferences\_guidelines = \{
     
        \hspace*{1.5em} 'include': '',
        
        \hspace*{1.5em} 'exclude': 'IMPORTANT - Omit ALL color information from descriptions. Do not mention colors of objects, clothing, environments, or any visual elements.'
    
    \}
\end{minipage}
}

\subsection{Customization Prompt: Description Emphasis}

\setlength{\fboxsep}{8pt} 
\setlength{\fboxrule}{0.5mm}
\vspace{0.2cm}
\noindent
\fcolorbox{black}{lightblue}{ 
    \begin{minipage}{0.94\linewidth}
    emphasis\_guidelines = 
    \{
    
    \hspace*{1.5em} 'character': 'Prioritize character-related details such as appearance, expressions, gestures, actions, and interactions. Focus on what people are doing and how they are doing it.',

    \hspace*{1.5em} 'environment': 'Prioritize spatial descriptions, atmosphere, setting, background elements, layout, lighting, and environmental textures. Focus on where the action takes place and the mood of the setting.',
        
    \hspace*{1.5em} 'general': 'Provide balanced descriptions following the general AD guidelines. Give equal attention to all visual elements.',
        
    \hspace*{1.5em} 'instructional': 'Prioritize the main plot or instructional content. Focus on plot progression, cause-effect relationships, and key narrative developments. Ensure descriptions and transitions between scenes are strongly tied to story or instructional continuity. Secondary visual details should be included only when they enhance plot understanding.'

\}
    
\end{minipage}
}

\subsection{Interactive Visual Question Answering Prompt}

\setlength{\fboxsep}{8pt} 
\setlength{\fboxrule}{0.5mm}
\vspace{0.2cm}
\noindent
\fcolorbox{black}{lightblue}{ 
    \begin{minipage}{0.94\linewidth}
You are a visual question answering aide for blind and low vision users.

Your task is to answer the question for the user in the context of the video and the provided screenshots.

You have to only answer related to the main frame, the adjacent frames are only for context.

Answer succinctly and naturally. Do not mention screenshots.

\{ -- QUESTION -- \}

\{ -- MAIN\_VIDEO\_FRAME -- \}

\{ -- ADJACENT\_FRAMES -- \}

\{ -- VIDEO\_AUDIO\_DESCRIPTIONS -- \}
\end{minipage}
}

\section{Codebook of VQA Types with Definitions and Examples}
\label{appendix:codebook}
\begin{table*}[ht]
\centering
\resizebox{0.98\linewidth}{!}{\begin{tabular}{p{3cm}p{6cm}p{6cm}}
\toprule
\textbf{Code Name} & \textbf{Code Description} & \textbf{Prompt Example} \\
\toprule
Describe Scene & Gathering visual information about the overall scene. &``Describe what is happening now'' \\

Identify Color & Questions or statements asking about describing colors of objects and subjects & ``What color is the thread?'' \\

Identify Presence & Determine whether an object, person, or entity is visible or exists in the frame/video. & ``How many people are gathered? \\
Identify Subject & Identifying what something is or represents. & ``What kind of cheese is this?'' \\
Identify Feature & Identifying features of a subject, such as size, clothing, and type. & ``What size is the solid state drive?'' \\
Describe Character & Describing people or characters' appearances. & ``How do the protestant and Orthodox priests look different?'' \\
Infer from video & Questions that refer to events, objects, dialogue, or information that may occur elsewhere in the video, rather than being answerable solely from the current video second. & ``Who said `I'm not on your team'?'' \\
\bottomrule
\end{tabular}}
\end{table*}

\end{document}